\PassOptionsToPackage{unicode}{hyperref}
\PassOptionsToPackage{hyphens}{url}
\documentclass[
]{article}
\usepackage{amsmath,amssymb}
\usepackage{lmodern}
\usepackage{iftex}
\ifPDFTeX
  \usepackage[T1]{fontenc}
  \usepackage[utf8]{inputenc}
  \usepackage{textcomp} 
\else 
  \usepackage{unicode-math}
  \defaultfontfeatures{Scale=MatchLowercase}
  \defaultfontfeatures[\rmfamily]{Ligatures=TeX,Scale=1}
\fi
\IfFileExists{upquote.sty}{\usepackage{upquote}}{}
\IfFileExists{microtype.sty}{
  \usepackage[]{microtype}
  \UseMicrotypeSet[protrusion]{basicmath} 
}{}
\makeatletter
\@ifundefined{KOMAClassName}{
  \IfFileExists{parskip.sty}{%
    \usepackage{parskip}
  }{
    \setlength{\parindent}{0pt}
    \setlength{\parskip}{6pt plus 2pt minus 1pt}}
}{
  \KOMAoptions{parskip=half}}
\makeatother
\usepackage{xcolor}
\usepackage[margin=1in]{geometry}
\usepackage{longtable,booktabs,array}
\usepackage{calc} 
\usepackage{etoolbox}
\makeatletter
\patchcmd\longtable{\par}{\if@noskipsec\mbox{}\fi\par}{}{}
\makeatother
\IfFileExists{footnotehyper.sty}{\usepackage{footnotehyper}}{\usepackage{footnote}}
\makesavenoteenv{longtable}
\usepackage{graphicx}
\makeatletter
\def\maxwidth{\ifdim\Gin@nat@width>\linewidth\linewidth\else\Gin@nat@width\fi}
\def\maxheight{\ifdim\Gin@nat@height>\textheight\textheight\else\Gin@nat@height\fi}
\makeatother
\setkeys{Gin}{width=\maxwidth,height=\maxheight,keepaspectratio}
\makeatletter
\def\fps@figure{htbp}
\makeatother
\providecommand{\tightlist}{%
  \setlength{\itemsep}{0pt}\setlength{\parskip}{0pt}}
\newlength{\cslhangindent}
\newlength{\csllabelwidth}
\newlength{\cslentryspacingunit} 
\newenvironment{CSLReferences}[2] 
 {
  \setlength{\parindent}{0pt}
  \ifodd #1
  \let\oldpar\par
  \def\par{\hangindent=\cslhangindent\oldpar}
  \fi
  \setlength{\parskip}{#2\cslentryspacingunit}
 }%
 {}
\usepackage{calc}

\usepackage{float} 
\ifLuaTeX
  \usepackage{selnolig}  
\fi
\IfFileExists{bookmark.sty}{\usepackage{bookmark}}{\usepackage{hyperref}}
\IfFileExists{xurl.sty}{\usepackage{xurl}}{} 
\hypersetup{
  pdftitle={What is a good doge? Analyzing the patrician social network of the Republic of Venice},
  pdfauthor={JJ Merelo},
  hidelinks,
  pdfcreator={LaTeX via pandoc}}

\title{What is a good doge? Analyzing the patrician social network of
the Republic of Venice}
\author{JJ Merelo}
\date{2023-02-20}

\begin{document}
\maketitle
\begin{abstract}
The Venetian republic was one of the most successful trans-modern
states, surviving for a millennium through innovation, commercial
cunning, exploitation of colonies and legal stability. Part of the
success might be due to its government structure, a republic ruled by a
doge chosen among a relatively limited set of Venetian patrician
families. In this paper we analyze the structure of the social network
they formed through marriage, and how government was monopolized by a
relatively small set of families, the one that became patrician first.
\end{abstract}

\hypertarget{introduction}{%
\subsection{Introduction}\label{introduction}}

During one thousand years, the Venetian republic managed to raise from a
small fiefdom under suzerainty of the Byzantine empire to the biggest
empire in the Mediterranean, and then maintain its existence until the
\emph{Napoleonic} storm reconfigured the map of Europe for ever
(Horodowich 2013).

One of the main factors that influenced in that stability was the way it
chose its presidents, or \emph{doges}, as well as the constraints put on
what they could effectively do, as well as a clear separation of powers
that created a plethora of institutions with more or less clearly
defined functions (Cecchini and Pezzolo 2012). The mercantile
orientation of these institutions certainly helped, since it created a
safe and secure environment where people from all over the world could
do business. But it was as extremely rare for a doge to be deposed due
to corruption, or ousted by a coup d'ètat, as it was extremely common
all over the Italian peninsula and elsewhere.

However, Venice was a far cry from a democracy. Nobles
(\emph{patriziato}) and the people (\emph{popolani}) were totally
separate, with the patricians being the only one with a possibility of
occupying any job going from simple administrative jobs, to the highest
offices such as the Maggior Consiglio (equivalent to a Privy Council),
the Senate or the \emph{Consiglio dei dieci}, that worked as higher
instance tribunal as well as home office. This effectively implied that
a certain number of families ruled the Republic, from its inception to
the end. This rule was, also, increasingly authoritarian, with a tight
grip on power, that was, however, distributed among the different
institutions.

It is, then, difficult to understand how what caused upheaval and
instability in most Medieval and post-Medieval states instead was not a
factor leading to the demise of the institution of the Doges. This paper
tries to contribute to understand why, by studying the social network
formed by these patrician families, and how it is related to the number
of doges these families contributed as governors of the republic.

The rest of the paper has been organized as follows: coming up, a brief
revision of the state of the art. The extraction and preparation of data
is described in the \protect\hyperlink{dataset}{section that comes
next}. The resulting social network will be analyzed in Section
\protect\hyperlink{sn}{The Venetian patrician families social network}.
We will finally draw some conclusions and present future work in the
last section.

\hypertarget{state-of-the-art}{%
\subsection{State of the art}\label{state-of-the-art}}

Social network analysis has increasingly become a tool in the toolset of
historians, but it has not been used comprehensively; it is still
relatively rare to find this kind of studies. Most studies focus on
trade or mercantile networks, since these are the ones that have
relatively better coverage. The trading networks of Venice were studied
in (Apellániz 2013), focusing on how low-ranking or colonial individuals
used mercantile networks to penetrate other social circles, or even how
the manipulation of these helped mobility, social and otherwise, for
them. In this, the power of social network analysis to first analyze,
and then, by focusing on the position of specific nodes and
mesoanalysis, gather insights on social mechanisms, was revealed. Two
quantities were analyzed: betweenness and closeness centrality. The
first measure is related to the position of a node (person) in a
specific network, and how passing through that node is a necessity to
communicate different parts of the network. Closeness centrality, on the
other hand, expresses how difficult is, for a specific actor, to reach
any other part of the network. These two quantities are essential to
have a high-level understanding of emerging properties of the network,
such as modularization, division between different groups or cliques,
and also how network-correlated attributes, such as wealth, arise. Based
on their books, Ryabova (2019) analyses the social networks of a trading
family, the Soranzo, which are also one of the \emph{case vecchie},
please see below for a brief history of the Venetian nobility. This
papers explains how mercantile companies were usually temporal
partnerships in Venice, which augmented the importance of the social
network as such. The study of commercial partnerships for this firm
shows how extensive was commerce with German cities (possibly belonging
to the Hansa league), which was even bigger than the one with Venice
itself. Although there's no attempt to analyze the ego network beyond
the connections between the center and others, it shows many patrician
families among the mercantile partners, revealing how ties between
nobles occurred at many different levels, political, familiar, and
mercantile.

Other approaches to the study of doges, however, are possible. (Smith,
Crowley, and Leguizamon 2021) recently publish a paper studying how the
age of doges was chosen in such a way that, even if it was a lifelong
post, their terms were naturally limited. As a matter of fact, we have
computed the median number of years ruling to be
\texttt{r\ median(ages\$Reigning)}, with average at 9.3625.

We will analyze in this paper the social network of patrician families
revealed by the marriage of doges, as a way to understand the power
dynamics among different noble families, and thus have some insights on
the causes of the stability and longevity of the Venetian state. But let
us first present a brief history of these patrician families, and their
attributes.

\hypertarget{a-brief-history-of-venetian-patriciate}{%
\subsection{A brief history of Venetian
patriciate}\label{a-brief-history-of-venetian-patriciate}}

The Venetian patriciate has been always attached to the access to power
symbolized by the office of the doge. However, it has evolved with the
institution and the different instituted laws that constrained it.

The beginning of the republic of Venice (Madden 2012) is (possibly
mythically) set in 697, with Paolo Lucio Anafesto converting from a
Byzantine governor to the first leader of the independent city of
Venice. This tradition does not have any source, but at any rate, it
started the institution of doges as \emph{elected} leaders. Since back
then Venice was little more than a settlement, a popular assembly
elected them directly.

After a brief period when Venice was part of the exarchate of Ravenna
and was governed by a military administration, dogarate was restored in
742, with doges, once again, elected popularly. This period was not
particularly stable. Many doges were deposed or directly killed. In one
of such cases, Pietro IV Candiano and his son were locked in palace by
the populace, after which it was set afire. However, the first
restricting law was introduced in 1032: the doge was forbidden from
appointing a consul, and even more so, a privy council was appointed to
take care of enforcing that law; a council of the wise was created in
the next century to further control and counter-balance the executive
power of the doge.

But one of the key laws in the future governance of the Republic was the
\emph{Serrata} or ``closure'' that was introduced in the XIII century.
This law effectively \emph{closed} entering the Supreme Council
(\emph{Maggior Consiglio}) to most families, and consecrated the
creation of a group of patrician families. Although doges themselves
could be chosen from anywhere, it was usually after a life of serving in
different institutions, including this Maggior Consiglio, when they
became effectively known and eligible, so the highest seat was
effectively closed to a group of families. For some time, only 25
families, thereafter called \emph{vecchie}, were eligible for these
jobs; out of those families, 12 were called \emph{apostoliche}, and were
supposed to be among the founding families of the state; another four
were denominated \emph{evangelisti}: Giustinian, Corner, Bragadin and
Bembo. These families will, effectively, form the initial pool where the
doges were drawn from.

There were other sets of families that eventually became noble in time.
The \emph{nuove} or new were incorporated in the XIII and XIV century,
after the war against Genoa and the fall of Constantinople. Those
incorporated in the XIV century were called \emph{nuovissime}, or ``very
new''; there were also some families from ``abroad'' (mainly, the
Adriatic colonies placed where Croatia and Montenegro are now, and even
other cities in \emph{terraferma}, or the continent, such as Verona or
Parma). With the decadence of the Republic in the XVII century, you
could simply pay your way into the nobility. More than a hundred
families accessed that way, some in the very last years of the Republic.

This history of the nobility implied that it was not, by any extent of
the term, an uniform caste. Besides the age of their \emph{case} or
dynasties, there was another axis, which was simply wealth. The
\emph{barnaboti} (Zennaro 2018) where simply poor noble people, who had
nothing going for them except their name. Their appellation alluded to
the fact that they resided in a destitute residence for nobles in the
area of Saint Barnabas. We will need to see how families belonging to
the different layers of accession to nobility are placed within the
social network, and how this history has an influence on, or at least
correlation, with it.

\hypertarget{dataset}{%
\subsection{Dataset generation}\label{dataset}}

It is theoretically possible, but out of reach for people that are not
scholars, to access the whole marriage registry in the city of Venice.
It is relatively simple, however, to find the list of all doges and all
the women they were married to. Initially, I thought it was possible to
use a simple Wikidata query to do so, but unfortunately those Wikidata
predicates have not been populated yet. So the dataset had to be built
pretty much by hand. In general, researching Medieval (and later) social
networks is an endeavor complicated by the lack of records (Ryabova
2016), or the need to enter manually whatever data is gathered into a
spreadsheet for their analysis.

This is why we needed to create a data set specifically for this line of
research. The list of all doges, and the list of all \emph{dogaresse},
or doge's wives, has been extracted from Wikipedia pages, the list of
doges from the Spanish page (since it was already formatted in a list)
and the list of dogaresse from the Italian Wikipedia. These lists
included dates when they were dogaresse; they were matched to their
husbands by these dates. Couples were then double-checked, when some
possible doubt (due to dates) was found, using primary sources such as
(Cicogna and Nani 1863). In some cases, dates had to be fixed in the
original list, as is shown
\href{https://it.wikipedia.org/w/index.php?title=Dogaressa\&type=revision\&diff=128805136\&oldid=127173404}{in
this diff}. The fact that Venetian wives did keep their family names, or
at least were recorded that way, helps to identify them as members of a
specific family.

All in all, there were 129 doges through all history. Out of all of
these, 49 did not marry, or their marriage has not been registered. Out
of these couples, there were a few doges that married twice; most
married only once. Additional filtering and processing was done:

\begin{itemize}
\tightlist
\item
  Couples marrying into the same family were eliminated. Patrician
  families were quite extensive, and sometimes had different branches.
  it was not common, but it happened, that some doge was married to a
  dogaressa belonging to another branch of the family. For instance,
  Pietro III Candiano as well as Pietro IV Candiano married to women of
  the Candiano family; Domenico Selvo married in 1075 to Theodora Selvo;
  Ordelafo Falier married to Matelda Falier, and right after that one,
  Domenico Michele married to Alicia Michele.
\item
  The patrician family was identified by following a regular expression
  on the original data. The surname of the doge was registered as the
  last name before the date of rule, the dogaresse was the last
  expression in the string. However, the extracted surname did not
  always correspond to a patrician family. In some cases, mainly in the
  initial stages of the republic, only first names were known. In some
  other case, mainly in the case of dogaresse, only their country or
  land of origin was noted: for instance, Valdrada di Sicilia or Loicia
  da Prata were married to doges in the late XIII century.
\item
  Family names were canonicalized. Many patrician family names have two
  forms: the one in Veneto, the language spoken in Venice, and the one
  in Italian. For instance, Cornaro and Corner, or Michele and Michel.
  In a case, ``Mastropiero'', it was also known as ``Malipiero'';
  Gradenigo could also be written as Gradenico. These were normalized to
  a single form, in Veneto, which is the one found in most primary
  sources.
\end{itemize}

After these filters, there were only 57 left. Out of these, there were
only 35 different families that became ``Serenest princes'' or doges, 36
that were \emph{dogaresse}. Some families had members in both roles: 14
of them, a very small percentage of the total families; on the other
hand, a total of 57 participated in any of them.

Out of all the families that contributed to the top most office in the
republic, there were 37 that did not marry, or their marriages were not
registered. Some relatively important families, like Bembo, Erizzo or
Michele, appear in this list. Five o them, including Bembo, do appear as
dogaresse. This means that out of the 65 families that became doges,
just 26 are not linked by marriage to their families, at least with this
small sample of links that is the list of married doges. Most of these
non-married families, except five of them, Partecipazio, Candiano,
Ziani, Galbaio and Teodato, had a single doge to their account. These
are special cases: Partecipazio were the first doges, when office was
still kind of inherited (the doge nominated a ``consul'', which
succeeded him). So they \emph{had} to be married; same goes for
Candiano, except these took office a bit later. Galbaio and Teodato were
not even properly doges, but more dukes or military leaders that took
office in the first millennium. Ziani was in fact married, except that
he was married to one such Cecilia, with no family recorded. So in fact
families with more than one doge in them did include at least one
marriage link with other patrician families; as a matter of fact, the
Bembo family, for instance, included only one doge, Giovanni Bembo, but
two dogaresse, Cornelia and Felicità.

Looking at this from another point of view, there were 37 families with
unmarried doges, or unregistered marriages, and 39 with married doges.
But 12 families have both doges who married and other who did not. These
families were: Candiano, Memmo, Dandolo, Ziani, Morosini, Mocenigo,
Malipiero, Loredan, Grimani, Donato, Venier, Contarini. This leaves only
25 who, being doges, do not have a registered marriage. This is a
minority out of the 64 that did.

We can draw two conclusions from this: being married was the prevalent
civil state of the doges, and the marriage registry is a good enough
tool to research the social network of the patrician families of which
they sprung.

The dataset was also enriched with additional data from the Wikipedia:
mainly, the type of family; that is, if a family was \emph{vecchie},
\emph{nuove} or one or the other half dozen denominations that families
received. This data will eventually prove essential to understand the
position in the social network.

All data sets have been released at the
\href{https://github.com/JJ/venice-patrician-social-network}{GitHub
repository} under a free license.

\hypertarget{sn}{%
\subsection{The Venetian patrician families social network}\label{sn}}

Let us have a general look at the different families and how they relate
to each other.

\begin{figure}
\centering
\includegraphics{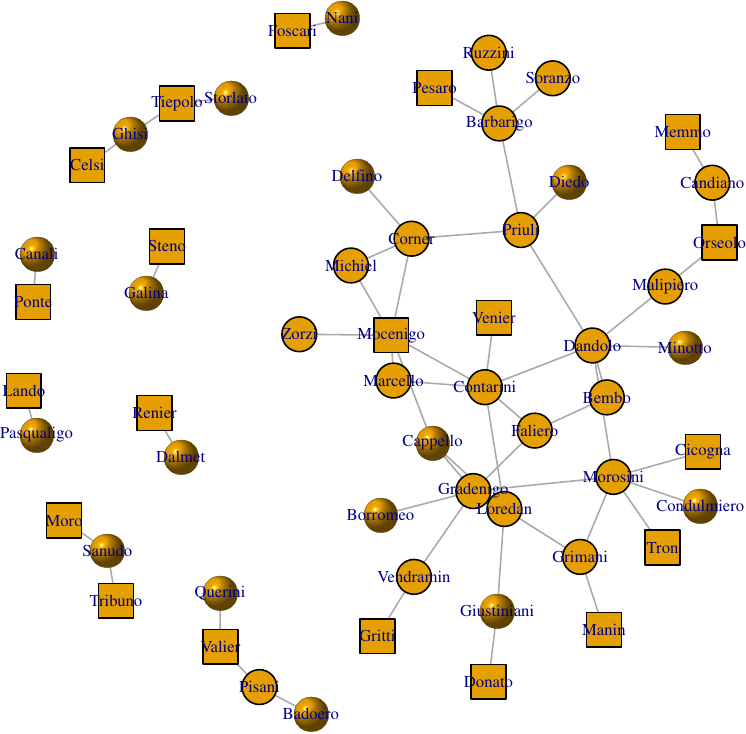}
\caption{Patrician families social networks through weddings. Circles
indicate that family has both; balls only doges, and squares only
dogaresse}
\end{figure}

This graph (in Figure 1) already reveals one of the features of this
social network: there is a connected core of different families, and
then small ``islands''. In general, these islands are composed of up to
four families that only managed to promote a member either to doge or
dogaressa; except for the one that includes Badoero, Pisani, Valier and
Querini, which includes a family, Pisani, that had both a doge (Alvise
Pisani, from 1735 to 1741, married to Elena Badoer) and a dogaressa
(Elisabetta Pisana or Pisani, from 1656 to 1658, married to Bertuccio
Valier). These are families that did not really make it, and accessed
power when the power of the Republic was already on the wane.

\begin{figure}
\centering
\includegraphics{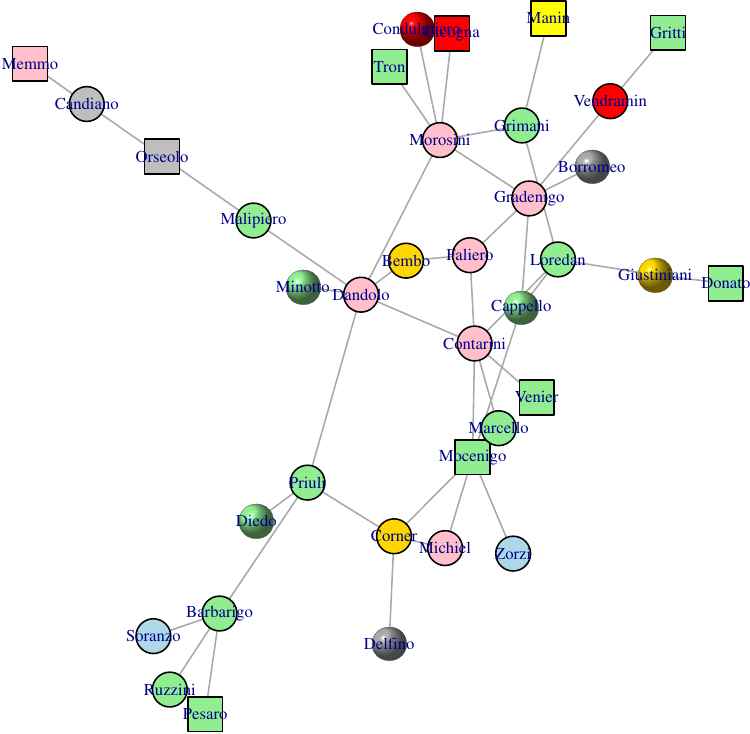}
\caption{Connected component of the social network, with network
coloring corresponding to the type of family. See text for
interpretation}
\end{figure}

In Figure 2 we show only the connected component, with color meaning as
follows: ``None''=``darkgray'', ``Ancient''=``lightgray'', ``Extinct
pre-serrata''=``gray'', ``Evangeliche''=``gold'',
``Nuove''=``lightgreen'',
``Nuovissime''=``red'',``Soldi''=``yellow'',``Vecchie''=``lightblue'',``Apostoliche''=``pink''.
The light-pink center of the graph, which has five \emph{apostoliche}
families, Contarini, Faliero, Gradenigo, Morosini and Dandolo, is also
``seeded'' with an \emph{evangeliche} family, Bembo. Other
``evangelical'' and ``apostolic'' families are in the periphery, but the
majority of the graph is occupied by the \emph{nuove} families (in light
green). Just a few of the \emph{nuovissime} are present here:
Condulmiero, Cicogna and Vendramin; and only one of the families that
bought their way into the patriciate, Manin, is included. This was,
also, the last doge before the fall of the republic. Other
\emph{vecchie} families, like Soranzo and Zorzi, are also present
peripherally.

The chart also reveals a certain grouping. Although factions were not so
much a factor in Venice politics, it can still show some insight into
the time-wise organization of weddings and how different groups were
formed through history; that is, it will tell us more about the temporal
organization of the social network than about antagonist groups in the
Senate or other institutions.

\begin{figure}
\centering
\includegraphics{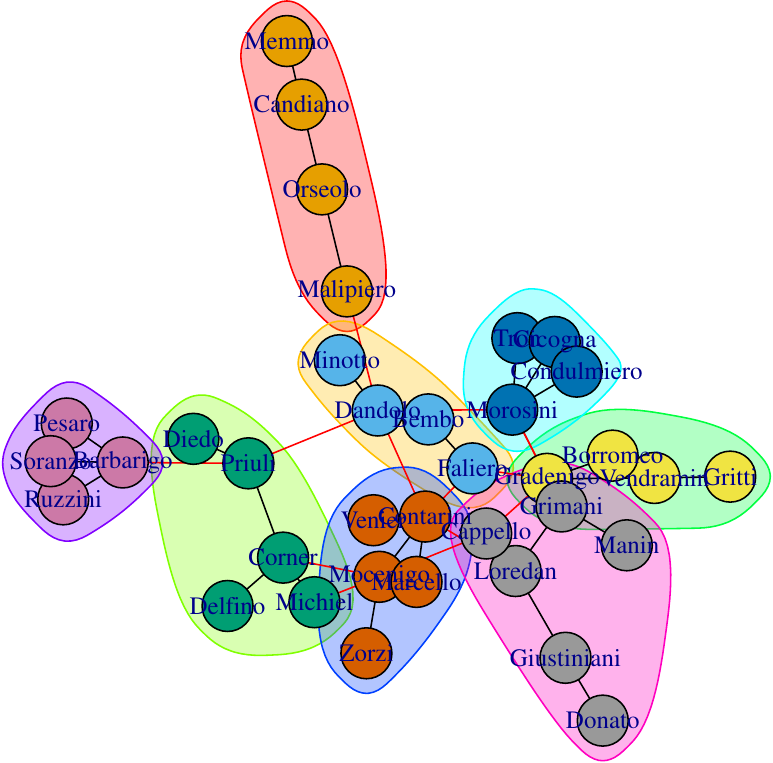}
\caption{Social groups in the central component of the patrician social
network; colors indicate different groups}
\end{figure}

Figure 3 shows seven different groups, every one of them with at most 6
elements. A group includes the Soranzo family, studied in their
commercial aspect in (Ryabova 2016). It is, as it can be seen, a
peripheral family only connected to the main component through the
Barbarigo family. But it is interesting to note that, despite that
family not showing up in the tables published by Ryabova above, we can
see the Pesaro (or Pexaro) family there, so it is plausible that
familial ties (with Barbarigo) led to commercial ties (with Pesaro) or
the other way round. As a matter of fact, it's entirely possible that
those ties exist too, only not at the doge level.

Another ``linear'' group includes the Orseolo, Candiano, Memmo and
Malipiero families; they are mostly ancient families, that did not
really survive beyond the era of elected doges. But it is also
interesting to note, in red, those links that bind families from
different groups, and how there is a specific family that includes them.
The Morosini and Contarini families, for instances, are two of those;
these are at the same time some of the oldest, \emph{apostoliche}
families, and they include lots of doges in them. Of course, that is the
reason why they have so many links, which are essentially weddings; but
the Corner or the Barbarigo family also do, and they don't have their
links places so strategically.

This visually observed reality can also be measured; some families seem
to be \emph{in the middle} of things. But how \emph{centered} are they?
So far we have been only visualizing the social network, let's analyze
the standing of the families in that social network numerically. In
general, betweenness centrality is one of the tools that measures that,
being a numerical representation of how \emph{needed} a node is to
communicate information between two separate areas of the network. This
is why it is one of the most widely used measurements in this area. For
patricians, being able to achieve commercial or political goals depended
on these family connections. A wedding, due to the fact that the bride's
family provided a dowry for her and that this dowry included in many
cases participation in business, was a link at many different levels. In
order to be induced into some entry level position (Chojnacki 1985)
indicates that sponsorship by your family and/or your bride's family
were essential for success. Since marriages were arranged, marrying
strategically, not only tactically, allowed families to achieve and then
keep their positions of powers in the Republic.

\begin{figure}
\centering
\includegraphics{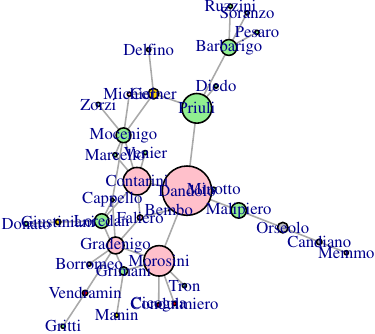}
\caption{Graph with node size proportional to betweenness}
\end{figure}

Figure 4 shows node size proportional to the family's betweenness, and
keeping the same coloring as Figure 2. It clearly shows that the Dandolo
family is the one with the highest betweenness, also highlighted by the
fact that it's at the center of the graph, since the layout algorithm
used (optimized by the \texttt{igraph} library (Csardi, Nepusz, et al.
2006)) takes that fact into account. Surrounding it, the Morosini,
Contarini and, surprisingly, Priuli family, the only one of the four
that is not an \emph{apostoliche}, but a \emph{nuova} family. This is a
curious case indeed, and shows the power of familiar ties in the
Venetian power structures. Priuli, except for the hereditary cases, is
the only family that spawned two consecutive doges: Lorenzo and Girolamo
Priuli. But more important than that is the marriages of the family, a
real who is who of the patrician families: Corner, Barbarigo\ldots{} And
Dandolo.

\begin{longtable}[]{@{}lr@{}}
\caption{Ranking of families according to betweenness}\tabularnewline
\toprule()
Family & Betweenness \\
\midrule()
\endfirsthead
\toprule()
Family & Betweenness \\
\midrule()
\endhead
Dandolo & 308.06667 \\
Morosini & 190.00000 \\
Priuli & 185.33333 \\
Contarini & 171.63333 \\
Gradenigo & 105.53333 \\
Barbarigo & 99.00000 \\
Malipiero & 96.00000 \\
Mocenigo & 92.40000 \\
Loredan & 90.40000 \\
Corner & 66.33333 \\
Orseolo & 66.00000 \\
Grimani & 52.00000 \\
Vendramin & 34.00000 \\
Candiano & 34.00000 \\
Giustinian & 34.00000 \\
Faliero & 33.60000 \\
Cappello & 24.50000 \\
Bembo & 6.00000 \\
\bottomrule()
\end{longtable}

This table shows the ranking of families according to betweenness, with
the Dandolo family on top. This family (Savoy 2015), known mainly after
Enrico Dandolo (Madden 2006), who personally led the IV crusade against
Constantinople, did not yield so many doges, and was rather at the end
of its tenure by the end of the Middle Ages. however, figure 3 reveals
how it became the link among different ``clans'' or cliques: the
``ancient'' one, through a marriage with the Malipiero family, the
``nuove'' through a marriage with the Priuli family, and the
``apostoliche'' through marriages into the Morosini and Contarini
families. In fact, these three families come next in the ranking; of
course the aforementioned families, in turn, leverage the connection
with the Dandolo family to uphold their positions of power. Since the
Dandolo family, one of the most famous throughout the history of Venice,
vanished when it started to decline, we can rather say that it was used
as instrument by the other families, rather than the other way round,
since the family had their heyday in the XIII century. The last Dandolo
to appear in the golden list was Zilia Dandolo, who married the Priuli
doge in the XVI century. At the same time, since the family is connected
to the ``ancient'' cluster, but only through a late marriage; the first
recorded marriage is the one between Enrico Dandolo and Felicità Bembo
that belonged to an ``evangelical'' family.

It should be noted, also, that these ``apostolic'' families occupy the
top posts in the betweenness ranking, accompanied by the ``new'' (after
the XIII century) families. Some of these families, like the Mocenigo
family, only appear in this connected graph as doges. At any rate, it
seems that the strategy of the families was for the evangelical families
to marry each other, as well as apostolic families, and for the new
families, to marry into the apostolic. This brought something like a
``dynastic'' succession, with new families taking over positions of
power after the XV century. However, new families acceding after those
did not really make it. It is likely that this lack of renewal in the
government, as well as the general decline in trade, technology and
production, contributed to the eventual demise of the republic.

\hypertarget{conclusions}{%
\subsection{Conclusions}\label{conclusions}}

The complicated way in which the doges were elected, and how individuals
were chosen to enter a career as officers of the Republic, guaranteed
that extremely wide, and variable, alliances had to be put in play.
These alliances were solidified through marriages, which integrated
politically as well as commercially the families of the two (romantic
and commercial) partners.

In this paper we have drawn two conclusions from the study of the social
network. One, that \emph{old} (\emph{vecchie}) families initially had a
firm grasp on power, mainly through the ones called \emph{apostoliche},
with \emph{evangeliche} and non-denominated old families coming next;
this power was conserved through familial and commercial alliances that
were kept in places for centuries. Some kind of dynastic renovation took
place with these being progressively substituted by the \emph{nuove}
families, but they arrived at power positions only when they created
alliances first with the \emph{vecchie}, mainly \emph{apostoliche},
families. The fact that this take over took place only after centuries,
with no violent changes, and that they were always pushed by ample
majorities, guaranteed the stability of the government, that was a
constant until the very end of the republic.

Two, that the position in the social network was highly correlated with
their position in an hypothetical ranking of the number of offices held
by that family through the centuries. More than individual contributions
(which were anyway important to be able to marry into one of the best
families), how links between families had been forged and kept were the
ones that made families raise to the top job in the Republic over and
over again over the centuries.

This, again, shows that quantitative studies over samples of social
networks are an excellent tool for understanding political history of
different countries, and contribute to the explanation of the stability
of the Republic's government: by raising to power only persons in
families that were able to forge the widest alliances, \emph{popular}
will was satisfied (as a matter of fact, \emph{patrician} will, since
they were the ones that effectively had the power to vote), and this was
one of the factors that contributed to the stability way beyond the
zenith of the Venetian empire.

There are two possible lines of works that can be pursued from this
report, among many. One of them will try to expand the social network by
including other kind of ties that have been registered, mainly
commercial. For the time being, this social network includes only a
fraction of all the families that became nobles in Venice. The second
will look at a more causal analysis, and specific strategies that the
families would follow, and how successful they were.

All data and code used in this paper is available under a free license
at \href{https://github.com/JJ/venice-patrician-social-network}{its
GitHub repository}. It includes R and Raku code, as well as the
different files. This paper has been written in RMarkdown, and its
source includes all code used to generate charts and tables.

\hypertarget{acknowledgments}{%
\subsection{Acknowledgments}\label{acknowledgments}}

This paper has been supported in part by project and DemocratAI
PID2020-115570GB-C22.

\hypertarget{references}{%
\subsection*{References}\label{references}}
\addcontentsline{toc}{subsection}{References}

\hypertarget{refs}{}
\begin{CSLReferences}{1}{0}
\leavevmode\vadjust pre{\hypertarget{ref-apellaniz2013venetian}{}}%
Apellániz, Francisco. 2013. {``Venetian Trading Networks in the Medieval
Mediterranean.''} \emph{Journal of Interdisciplinary History} 44 (2):
157--79.

\leavevmode\vadjust pre{\hypertarget{ref-institutions}{}}%
Cecchini, Isabella, and Luciano Pezzolo. 2012. {``Merchants and
Institutions in Early-Modern Venice.''} \emph{The Journal of European
Economic History} 41 (2): 89--114.
\url{https://www.proquest.com/scholarly-journals/merchants-institutions-early-modern-venice/docview/1778650464/se-2}.

\leavevmode\vadjust pre{\hypertarget{ref-chojnacki_1985}{}}%
Chojnacki, Stanley. 1985. {``Kinship Ties and Young Patricians in
Fifteenth-Century Venice.''} \emph{Renaissance Quarterly} 38 (2):
240--70. \url{https://doi.org/10.2307/2861664}.

\leavevmode\vadjust pre{\hypertarget{ref-cicogna1863storia}{}}%
Cicogna, Emmanuele Antonio, and Antonio Nani. 1863. \emph{Storia Dei
Dogi Di Venezia}. G. Grimaldo.

\leavevmode\vadjust pre{\hypertarget{ref-csardi2006igraph}{}}%
Csardi, Gabor, Tamas Nepusz, et al. 2006. {``The \texttt{igraph}
Software Package for Complex Network Research.''} \emph{InterJournal,
Complex Systems} 1695 (5): 1--9.

\leavevmode\vadjust pre{\hypertarget{ref-horodowich2013brief}{}}%
Horodowich, Elizabeth. 2013. \emph{A Brief History of Venice}. Hachette
UK.

\leavevmode\vadjust pre{\hypertarget{ref-madden2006enrico}{}}%
Madden, Thomas F. 2006. \emph{Enrico Dandolo and the Rise of Venice}.
JHU Press.

\leavevmode\vadjust pre{\hypertarget{ref-madden2012venice}{}}%
---------. 2012. \emph{Venice: A New History}. Penguin.

\leavevmode\vadjust pre{\hypertarget{ref-ryabova2016social}{}}%
Ryabova, Maria. 2016. {``The Social Network of the Soranzo Brothers as
Reflected by Their Account Books.''} \emph{Istoriya} 7 (6 (50)).

\leavevmode\vadjust pre{\hypertarget{ref-ryabova2019venetian}{}}%
---------. 2019. {``Venetian Trading Firm of the Soranzo Brothers
(1406-1434) and Its Commercial Network.''} \emph{Venetian Trading Firm
of the Soranzo Brothers (1406-1434) and Its Commercial Network},
229--53.

\leavevmode\vadjust pre{\hypertarget{ref-dandolomyth}{}}%
Savoy, Daniel. 2015. {``Keeping the Myth Alive: Andrea Dandolo and the
Preservation of Justice at the Palazzo Ducale in Venice.''}
\emph{Artibus Et Historiae}, no. 71: 9--29.
\url{https://www.proquest.com/scholarly-journals/keeping-myth-alive-andrea-dandolo-preservation/docview/1764752196/se-2}.

\leavevmode\vadjust pre{\hypertarget{ref-smith2021long}{}}%
Smith, Daniel J, George R Crowley, and J Sebastian Leguizamon. 2021.
{``Long Live the Doge? Death as a Term Limit on Venetian Chief
Executives.''} \emph{Public Choice} 188 (3): 333--59.

\leavevmode\vadjust pre{\hypertarget{ref-zennaro2018ritratto}{}}%
Zennaro, Nicolò. 2018. {``Un Ritratto Della Societ{à} Veneziana.''}

\end{CSLReferences}

\end{document}